\documentclass[aps,prl,10pt,twocolumn,showpacs,superscriptaddress]{revtex4-1}

\usepackage{amsmath}
\usepackage{latexsym}
\usepackage{amssymb}
\usepackage{bm}
\usepackage{graphics,epstopdf}
\usepackage{color}
\usepackage{hyperref}

\usepackage{newlfont}
\usepackage{amsfonts}
\usepackage{amsthm}
\usepackage{graphicx}
\usepackage{epsfig}

\newcommand{\ket}[1]{|{#1}\rangle}
\newcommand{\bra}[1]{\langle{#1}|}
\newcommand{\braket}[2]{\langle {#1} | {#2} \rangle}

\usepackage{times}

\usepackage[up]{subfigure}

\newcommand{\be}{\begin{equation}}
\newcommand{\ee}{\end{equation}}
\newcommand{\bc}{\begin{center}}
\newcommand{\ec}{\end{center}}
\newcommand{\bea}{\begin{eqnarray}}
\newcommand{\eea}{\end{eqnarray}}
\newcommand{\ba}{\begin{array}}
\newcommand{\ea}{\end{array}}

\begin{document}
\title{Super Quantum Search Algorithm with Weak Value Amplification and Postselection}

\author{Arun Kumar Pati}
\email{akpati@hri.res.in}
\affiliation{Quantum Information and Computation Group,\\
Harish-Chandra Research Institute, HBNI, Chhatnag Road, Jhunsi,
Allahabad 211 019, India}

\date{\today}

\begin{abstract}
We propose a new model of quantum computation which aims to speed up quantum algorithms assisted by the weak value amplification and ancillay quantum 
register with the pre- and postelection. Within this model, we show that a quantum computer can solve a data base search
of $N$ entries in one step with probability close to one for large $N$, provided the post-selection on the ancillary quantum register is successful.
In this model, to search a database of $N$ entries, the number of qubits grows from $n$ to $2n$, but there is a huge reduction in time
complexity. Physically, this can be understood as the effect of weak value amplification that arises due to the pre- and postselection of the ancillary register 
which interacts with the $n$ qubit register where quantum search is performed. This effectively accelerates the computation
and takes the state of quantum computer much ahead in time, compared to what one would obtain without weak value amplification and post-selection.
\end{abstract}

\maketitle

{\it  Introduction.--}
Quantum computers have the potential to solve problems that are beyond the reach of even the most powerful classical supercomputers. Feynman \cite{fey} has envisioned that the 
laws of quantum mechanics make it hard for classical computers to simulate the behavior of complex many-body quantum systems. This flared the interest that may be 
one should explore quantum computers to simulate quantum systems efficiently. After the pioneering idea by Deutsch \cite{dd}, we know that the superposition provides 
the massive parallelism and entanglement provides the fuel for quantum computers. Even though the race to build quantum
computers is already on the horizon and facing its own challenges, the ability to discover new 
quantum algorithms is quite demanding. 
In the standard model of quantum computation, we prepare 
the quantum computer in an $n$-qubit state which is an equal superposition of all possible logical states, apply the sequence of unitary transformations (quantum gates) depending 
on the algorithm to be implemented, 
and finally measurement is performed in the computational basis to retrieve the correct answer. This can also be phrased as `prepare-compute-read' model of quantum computer.
Over the last three decades, several quantum algorithms have been discovered \cite{ddj,shor,lov,lov1,vazi,sim,seth} that show the quantum advantage over classical computers.
Two notable algorithms are worth mentioning here, one is Shor's algorithm \cite{shor} that could efficiently factorize large numbers and the other is Grover's algorithm  \cite{lov} 
which aims to search in a fastest possible way a particular item from an unstructured database. Grover's algorithm can search a marked item in $O(\sqrt{N} ) $ steps for 
a database of $N$ entries and this quadratic speed-up has been proven to be optimal \cite{ben,boy,zal}. Though the 
importance of entanglement in Shor's algorithm \cite{noha}, Grover's algorithm \cite{slb} and general quantum algorithms have been \cite{jl} entrenched, it remains an open 
question as to how much entanglement is required, if at all, to achieve exponential
speed-up for quantum algorithm over classical algorithm.

{\em Weak value, pre- and postselection.--}
If a quantum system is pre-selected in a state and post-selected in an non-orthogonal state, then if that system is allowed to interact weakly with another 
quantum system, the state of the later system is affected by a strange value-called the `weak value' \cite{abl,aav}. Unlike the eigenvalues of a physical observable, 
the weak value can be complex in nature and arbitrarily large. The notion of two-time symmetric formalism \cite{abl} plays a major role in interpreting the weak value and  
the former has been exploited to propose quantum time-translation machine \cite{sandu} 
which can take a quantum system to future under suitable postselection.
The notion of modular value has been introduced which goes beyond the notion of weak value \cite{lev} and can arise even without weak interaction. 
Very recently, the concept of potent value 
has been defined which unifies the idea of weak value and modular value \cite{akp}. 
 This describes how a quantum system affects the state of the another system during
the time between two strong measurements corresponding to the pre- and post-selection.
The weak value amplification is a concept that has been 
extensively used in a myriad of applications in recent years.
It has been shown that weak  measurements can be used for interrogating quantum systems in a
coherent manner \cite{tol,cho}.
Among others, it plays important role in understanding the uncertainty principle in
the double-slit experiment \cite{wise1,mir}, resolving Hardy's paradox \cite{js},
analyzing tunneling time \cite{AMS951, AMS952}, and modifying the quantum mechanical decay law \cite{pdav}.
The optimal probe wavefunction and the upper bound for the amplification factor for the weak value has been explored in Ref.\cite{shika1,shika3}.
By expressing the wavefunction as a weak value of a projector, one can 
 measure the wavefunction of single photon \cite{jsl1,jsl2,jsl3}.
 Similarly, the advantage of weak value amplification in quantum
metrology has been investigated in details \cite{hof,todd,jord,just}.
Quite intriguingly, the notion of weak value also allows one to measure the average of any 
non-Hermitian operator in quantum mechanics \cite{pati} and this has been tested in interferometry without weak interaction and postselection \cite{sinha}.


{\it Quantum computing with weak value amplification and postselection.--}
In the circuitry model of quantum computation one considers only pre-selected quantum computer. 
Since quantum theory allows pre- and postselected ensembles as fundamental quantum ensembles and 
they contain more information than ensembles that are only
preselected, then a natural question is whether we can speed-up quantum algorithm assisted by quantum register which has been pre- and post-selected. 
The effect of postselection in quantum computing has been discussed earlier, where the postselection was supposed to be performed on the 
quantum computer itself and the role of ancillary quantum system with the pre- and postselection as a resource has not been exploited. 
Using the idea of postselection it has been argued 
that quantum computation can be carried out in
constant depth that cannot be simulated classically with high accuracy \cite{ter}. It has been argued that any problem that can be solved in BQP with polynomial 
quantum advice can also be solved in Probabilistic Poynomial-Time (PP) with polynomials-size classical advice \cite{scot}. Most importantly, it has been proved 
that postselection can be used to 
define a new complexity class, namely, Postselected Bounded-Error Quantum Polynomial-Time (PostBQP) with postBQP =PP  \cite{scot1}.
Using the relational time-symmetric framework a new perspective to some of the existing quantum algorithms have been presented \cite{eli}.
However, the effect of weak value amplification along with postselection have never been considered before in quantum computing. Specifically, we ask whether one can 
discover new quantum algorithms using the weak value amplification along with ancillary register in pre- and postselected states?

This motivates us to propose a new model of quantum computation which aims to speed up quantum search algorithm assisted by weak value amplification and ancillary quantum 
register with pre- and post-selected states. This model, for brevity, we name it as Quantum Computation assisted with weak value amplification and postselection (WVAP) or 
WVAP Quantum Computation. This requires two quantum registers each having $n$-qubits, where the first register is our quantum computer where the actual algorithm will be  
implemented and the second register is allowed to interact with the quantum computer with specific pre- and postselected states at intermediate times. 
In a sense, the pre- and postselected quantum register acts as a resource for quantum computation. 
We show that if we have access to the ancillay quantum register with the specific pre and post-selection, then 
a quantum computer can solve a database search of $N$ entries in one step with probability close to one for large $N$. In this model, we do not allow
postselection on the quantum computer itself where algorithm is being implemented.
Even though to search a database of $N$ entries the number of qubits double, there is a huge reduction in time
complexity. This exponential quantum advantage is the effect of weak value amplification that arises due to pre- and post-selection of $n$ qubit ancillary 
register which interacts with the quantum computer where quantum search is performed. Therefore, this may provide a new mechanism for accelerating the computation
which takes the state of quantum computer much ahead in time, compared to what one would obtain without weak value amplification and post-selection. Thinking differently, 
we can say that 
effect of weak value amplification and postselection creates the potent operator which takes 
the quantum computer to future (similar to time-translation machine), thus deciphering the item 
to be searched in one oracle query. It may be stressed that without the weak value amplification, exponential speed up may not be possible just by using postselection. 
Of course, the exponential speed-up does not come for free. There is a price we have to pay for the super search algorithm and that is fetched by the ancillary quantum register.

{\it Potent Value and potent operator.--}
For the sake of completeness, we briefly introduce the notion of potent value and potent operator here. When two systems interact weakly, we know that if one system is 
subject to the pre- and postselection, 
then the dynamics of the other system is affected by the weak value. If the coupling is not small, then how does the pre- and postselected system affect the other
system? To answer this question, we need to introduce the concept of potent value \cite{akp} which is again kind of `strange value' corresponding to observable $A$ that 
affects the state of another system for arbitrary coupling strength. 
Consider two quantum states $\ket{\Psi} \otimes \ket{\psi_i} \in {\cal H}_1 \otimes {\cal H}_2$.  Let these two quantum systems evolve under the action of an interaction Hamiltonian
as given by 
\begin{align}
 \ket{\Psi} \otimes \ket{\psi_i}  & \rightarrow  e^{-\frac{i}{\hbar}g O \otimes A} \ket{\Psi} \otimes \ket{\psi_i},
\end{align}
where $\ket{\Psi}$ is the initial state of the first system, $\ket{\psi_i}$ is the pre-selected state of second system, $O$ and $A$ are the observables of the first and second system, 
respectively.
This evolution can be expressed as 
\begin{align}
 \ket{\Psi} \otimes \ket{\psi_i}  & \rightarrow \sum_k \bra{k} e^{-\frac{i}{\hbar}g O \otimes A} \ket{\Psi} \ket{k} \otimes \ket{\psi_i} \nonumber\\
 &  = \sum_k  \ket{k} \otimes  A_k \ket{\psi_i},
\end{align}
where $A_k =  \bra{k} e^{-\frac{i}{\hbar}g O \otimes A} \ket{\Psi}$ and $\ket{k}$ is an orthonormal basis for the 
first system. The set of operators $A_k$ are in general are not unitary and act on the state of the second system in ${\cal H}_2$. 
The state of the first system, after the general interaction and upon post-selection of the system in the 
state $\ket{\psi_f}$, is given by 
\begin{align}
  \ket{\Psi_f} = {\cal N} \sum_k \frac{\bra{\psi_f} A_k \ket{\psi_i}}{ \braket{\psi_f}{\psi_i}} \ket{k},
\end{align}
where ${\cal N}$ is a normalization factor. The set of complex numbers defined below are called potent values that depend on the observable of the 
second system. These are given by
\begin{align}
\langle A^{(k)} \rangle_P  =  A^{(k)}_P (\psi_f|\psi_i) = \frac{\bra{\psi_f} A_k \ket{\psi_i}}{ \braket{\psi_f}{\psi_i}}.
\end{align}
Therefore, the final state of the first system is given by
\begin{align}
  \ket{\Psi_f} = {\cal N} \sum_k  \langle A^{(k)} \rangle_P  \ket{k},
\end{align}
where ${\cal N}^2= \frac{1}{ \sum_k  |\langle A^{(k)} \rangle_P |^2 } $.
Here, the effect of pre- and post-selected system on the first system is completely described by a set of potent values $\langle A^{(k)} \rangle_P$.
Since the final state of the system can also be expressed as $\ket{\Phi_f} = \sum_k C_k \ket{k} $ with where $C_k = {\cal N} \langle A^{(k)} \rangle_P  $,
the potent values of the second system actually describe the final 
state of the quantum system in ${\cal H}_1$ completely. Notice here that like the weak value, these set of potent values can be complex and arbitrarily large. Also, by measuring 
the potent values via the measurement of non-Hermitian operators \cite{pati}, we can determine the state of the quantum system. 

The potent values generalize and unify the notion of weak values \cite{aav} and modular values \cite{lev}. 
If $g << 1$, i.e., for the weak interaction, the potent values are related to weak values 
and for arbitrary interaction the potent values are related to modular values. In the weak coupling limit $g << 1$, these potent values 
result in the weak value effectively, thus leading to the final state 
 of the first system as given by 
 \begin{align}
 \ket{\Psi_f} = {\cal N}  e^{-\frac{i}{\hbar}g A_W(\psi_f|\psi_i)  O } \ket{\Psi},
\end{align}
where $A_W(\psi_f|\psi_i) = \frac{\bra{\psi_f} A_ \ket{\psi_i}}{ \braket{\psi_f}{\psi_i}}$ is the weak value of the observable $A$.

  When a quantum system interacts with a pre- and postselected quantum system, one can define another concept which we call as the potent operator.
 Consider a composite system consisting of two quantum systems that evolves under unitary evolution as given by
 $$\ket{\Psi} \otimes \ket{\psi} 
 \rightarrow U \ket{\Psi} \otimes \ket{\psi_i} = e^{-\frac{i}{\hbar}g O \otimes A} \ket{\Psi} \otimes \ket{\psi_i},$$
 where $g$ is an arbitrary coupling strength. If we postselect the second system in the state $\ket{\psi_f}$, 
 the final state of the first system, up to normalization, is described by
 \begin{align}
\ket{\Psi_f}  = \frac{\bra{\psi_f} e^{ - \frac{i}{\hbar} g O \otimes A}  \ket{\psi_i}}{ \braket{\psi_f}{\psi_i} } \ket{\Psi} =  U_P(\psi_f|\psi_i) \ket{\Psi},
\end{align}
where  
 \begin{align}
 U_P(\psi_f|\psi_i) =  \frac{\bra{\psi_f} e^{ - \frac{i}{\hbar} g O \otimes A}  \ket{\psi_i}}{ \braket{\psi_f}{\psi_i} } 
 \end{align}
  is the potent operator that acts on the first Hilbert space ${\cal H}_1$. 
  
The potent operator \cite{akp} has interesting connection to weak values \cite{aav} and modular values \cite{lev}. Suppose that the joint unitary operator is a 
conditional unitary operator 
 $U = \sum_k \Pi_k \otimes V_k$ where $\Pi_k$'s are
the projection operators acting on ${\cal H}_1$ with $\Pi_k \Pi_l = \Pi_k \delta_{kl}$ and $V_k$'s are the set of unitaries that act on ${\cal H}_2$. 
In this case the potent operator is given by
\begin{align}
 U_P(\psi_f|\psi_i) =  \sum_k \frac{\bra{\psi_f} V_k \ket{\psi_i}}{ \braket{\psi_f}{\psi_i}} \Pi_k =  \sum_k  \langle V_k \rangle_M \Pi_k ,
 \end{align}
where the potent operator depends on the set of modular values $\langle V_k \rangle_M $ in the second system as well as on the local projection operators 
in ${\cal H}_1$.
The final state of the first system, up to normalization, is given by
\begin{align}
  \ket{\Psi_f} =  \sum_k  \langle V_k \rangle_M  \Pi_k  \ket{\Psi}.
\end{align}
If the unitary $V_k$'s are also Hermitian operators, then $\langle V_k \rangle_M =  \langle V_k \rangle_W$ and they can represent weak values.
Thus, the notion of weak values and modular values can arise in a more general context without weak interaction and specific nature of systems.
 
{\it Super quantum search algorithm.--}
In quantum searching, we are given an unknown binary function $f_y(x)$, which returns $1$ for a
unique ‘target’ value $x = y$ and $0$ otherwise, where $x = 0, 1, 2, . . . , N-1 $, with $N = 2^n$ . Our goal is to
find $y$ such that $f_y(x) = 1$ for $x=y$. Here, $N$ items of the database are mapped onto the states of $n$ qubits. The quantum search 
problem is to reach the target item in a shortest possible time, thus amounts to maximizing the overlap between the target state and the 
final state of the quantum computer after algorithm has been implemented.

$\bullet$ New algorithm begins with two quantum registers each consisting of $n$ qubits, where $n$ is the number of
qubits necessary to represent the search space of size $2^n = N$. The first $n$-qubit register is the quantum computer where the item to be searched 
is stored and the second register is the ancillary system which is used for the pre- and postselection.

$\bullet$ Let the $n$-qubits of the quantum computer are initialized to $\ket{0}^{\otimes n}$ and the initial state of the ancillary system is pre-selected in the state 
$\ket{\psi_i}$ ( we will specify this later). 
Apply the Hadamard gate to each qubit in the first register. The state of the first register 
is $\ket{\Psi} = \frac{1}{\sqrt N} \sum_{x=0}^{N-1} \ket{x} = \cos \theta \ket{\bar{y}} + \sin \theta \ket{y}$, where  
$\ket{\bar{y}} = \frac{1}{\sqrt {N-1}} \sum_{x \not= y} \ket{x} $, $\ket{y}$ is the target state we wish to find and $\sin \theta = \frac{1}{\sqrt N}$.

The joint state of the quantum computer and the ancillary quantum register is given by

\begin{align}
\ket{\Psi} \otimes \ket{\psi_i} =  (\cos \theta \ket{\bar{y}} + \sin \theta \ket{y}) \otimes \ket{\psi_i}.
\end{align}

$\bullet$ We allow the quantum computer and the ancillary quantum register to interact by a quantum gate which is a conditional unitary operator $U$ given by 

\begin{align}
 U = {\bar \Pi_y}  \otimes I  + \Pi_y \otimes V,
\end{align}
where $ {\bar \Pi_y} = (I- \Pi_y)$, $\Pi_y = \ket{y}\bra{y}$ and $V$ is a unitary operator that acts on $n$ qubit ancillary register. In particular, 
$V = H^{\otimes n} \sigma_x^{\otimes n} H^{\otimes n} I_w$, where $H$ is the Hadamard gate, $\sigma_x$ is the $X$-gate and $I_w = (I - 2 \ket{w} \bra{w}) $
is the reflection operator which is Hermitian and self-inverse  with $\ket{w}$ being any one of the orthonormal state of the second register in the computational basis. 
The unitary $V$ can also be expressed as $V = \sigma_z^{\otimes n}  I_w $ as $H \sigma_x H = \sigma_z$.
If the quantum computer (first $n$-qubit register) is in the state $\ket{y}$, then the unitary operator $V$ is applied on the ancillary register
and does nothing if the quantum computer is in any other state different than $\ket{y}$.
Note that to implement $U$ we need only one oracle query on $n$ qubit quantum computer that contains the search item (see Eq(20)).

The unitary operator allows two registers to evolve as
\begin{align}
 \ket{\Psi} \otimes \ket{\psi_i} \rightarrow    {\bar \Pi_y} \ket{\Psi} \otimes \ket{\psi_i} 
  + \Pi_y  \ket{\Psi} \otimes V \ket{\psi_i}.
\end{align}

$\bullet$  After this unitary operator, we post-select the ancillary register in the state $\ket{\psi_f}$. If post-selection is successful, then 
the final state of the quantum computer that contains the item to be searched  is given by 
\begin{align}
\ket{\Psi_f}  =  {\cal N} [\cos \theta \ket{ \bar{y} } + \sin \theta  \langle A \rangle_W \ket{y},
\end{align}
where $ \langle V \rangle_W = \frac{\bra{\psi_f} V \ket{\psi_i}}{ \braket{\psi_f}{\psi_i} }  $ is called as the  
weak value of $V$ in the pre-selected state $\ket{\psi_i}$ and postselected state $\ket{\psi_f}$  and ${\cal N}$ is the normalization factor with ${\cal N}^2 = 
\frac{1}{\cos^2 \theta + \sin^2 \theta  |\langle A \rangle_W|^2}$. It may be mentioned that 
$ \langle V \rangle_W$ can also be called as the modular value $ \langle V \rangle_M$ as $V$ is a unitary operator. 
As we will see later, $V$ is not only unitary but also Hermitian, therefore, we can call it as the weak value of $V$. We can also view the 
final state of the quantum computer (14) as a result of the action of the potent operator 
$U_P(\psi_f|\psi_i) =  {\bar \Pi_y}  + {\Pi_y} \langle V \rangle_W$, i.e., $\ket{\Psi_f} = {\cal N} U_P(\psi_f|\psi_i) \ket{\Psi}$.

$\bullet$
Now, we ask what is the probability of finding the target state $\ket{y}$ when quantum computer is in the state $\ket{\Psi_f}$ ? This is given by 
\begin{align}
p = |\braket{y}{\Psi_f}|^2 = \frac{ \sin^2 \theta  |\langle A \rangle_W |^2 }{ \cos^2 \theta + \sin^2 \theta  |\langle A \rangle_W |^2 }.
\end{align} 


Can the weak value amplification help the search algorithm beyond what is allowed by the Grover algorithm? The answer is yes.
By suitable choice of pre- and postselected states for the ancillary quantum register and the unitary operator $V$ acting on the ancillary register, 
one can show that the probability
of finding the target state can actually be close to one, for large $N$.

   Note that we define the pre-selected state for the second register as the one which is the state of the ancillary system just before we apply the 
   unitary interaction between these two quantum 
   registers. We could, in principle, pre-process the $n$ qubit ancillary register state as we wish.
Consider the ancillary state of quantum register pre-selected in $\ket{\psi_i} =  I_w H^{\otimes n} \ket{0}^{\otimes n} = I_w \ket{\psi}$, where
$\ket{\psi} =  \frac{1}{\sqrt N} \sum_{x=0}^{N-1} \ket{x} = \cos \theta \ket{\bar{w}} + \sin \theta \ket{w}$, with  
$\ket{\bar{w}} = \frac{1}{\sqrt {N-1}} \sum_{x \not= w} \ket{x} $ and $\ket{w}$ is any one of the orthonormal basis of $n$ qubit 
ancillary register, and $I_w = (I - 2 \ket{w} \bra{w}) $. Therefore, the pre-selected state of the ancillary register is given by 
\begin{align}
\ket{\psi_i}  =  \ket{ \psi } - \frac{2}{\sqrt N} \ket{w}.
\end{align}

Next, we perform measurement on the second quantum register with the post-selected state
$\ket{\psi_f} =  \ket{ -}^{\otimes n}$, where $\ket{ -} = H \ket{1}$. Using 
$\braket{\psi_f}{\psi} = 0$ and $\braket{\psi_f}{w} = \pm \frac{1}{\sqrt N}$, we have 

\begin{align}
|\braket{\psi_f}{\psi_i}|^2    = \frac{4}{ N^2}.
\end{align}

Now, with the choice of $V = H^{\otimes n} \sigma_x^{\otimes n} H^{\otimes n} I_w$, we can check that $|\bra{\psi_f}V \ket{\psi_i}|^2    = 1$.
With this choice of the pre- and post-selection, we have the modulus square of the weak value modulus as  $|\langle A \rangle_W|^2 = \frac{N^2}{4}$. 
Therefore, the probability of finding the target state,
upon post-selection on the ancillary register, is given by 

\begin{align}
p    = \frac{N^2}{ N^2 + 4(N-1)}.
\end{align}
Hence, for the large database search, we have $ p= 1 - O(\frac{1}{N})$, i.e., the probability of finding the correct item approaches one. This completes 
the proof of the main result.

Now, we discuss how many oracle queries are required to implement $U$. Below, we will show that to implement $U$ we need only one oracle query. 
An oracle is a black-box function that evaluates a function to check the desired solution. The quantum oracle is a quantum black-box which can 
recognize if the system is in the correct state. If the system is indeed in the correct state,
then the oracle will give $f_y(x) = 0$ for $x \not= y$ and $f_y(x) =1$ for $x=y$. In the Grover search algorithm, the quantum oracle implements 
the unitary transformation as given by
\begin{align}
 \ket{x} \ket{j} \rightarrow  \ket{x} \ket{j \oplus f_y(x) },
\end{align}
where $\ket{x}$ is an $n$-qubit state of the quantum computer and $ \ket{j }$ is a single qubit state. 
Quantum oracle can also be expressed as 
$\ket{x} \rightarrow (-1)^{f_y(x)} \ket{x}$, 
where $f_y(x) = 1$ if $x$ is the correct state, and $f_y(x) = 0$ otherwise. This can be represented as a unitary operator 
$I_y = (I -2 \ket{y} \bra{y} )$ so that $I_y \ket{y} = - \ket{y}$ and $I_y \ket{x} = \ket{x}$ if $x \not= y$.
Note that in the case of Grover search, we need to apply the Grover operator $G = - I_0 I_y$ repeatedly $O(\sqrt{N})$ times to find the target state, where 
$I_0 = (I- 2 \ket{\Psi} \bra{\Psi})$. In the case of super quantum search, the unitary operator $U$ given in (2) can be expressed as 
\begin{align}
 U = I \otimes P_1  + I_y \otimes P_2,
\end{align}
 where $P_1  =  \frac{1}{2} (I + V)$ and $P_2 = \frac{1}{2} (I - V)$. 
 Notice that $V = \sigma_z^{\otimes n} I_w$ and we can always chose $\ket{w}$ as an eigenstate of  $ \sigma_z^{\otimes n} $ with eigenvalue $+1$, i.e., 
 $ \sigma_z^{\otimes n} \ket{w} = \ket{w}$. In this case, we can check that $ \sigma_z^{\otimes n} $ and $I_w$ commute and $V$ is not only unitary but also a Hermitian
 operator.  Furthermore, $P_1  =  \frac{1}{2} (I + V)$ and $P_2 = \frac{1}{2} (I - V)$ 
 becomes two orthogonal projectors on the $n$-qubit ancillary register, i.e., $P_i^2 =P_i (i= 1,2)$ and $P_1P_2 =0$ .
 This follows from the fact that $V$ is unitary and Hermitian. Thus, the unitary operator given in (2) for super quantum search algorithm is a 
 controlled Grover oracle query that uses the function evaluation only once. Also, notice that there is an interesting symmetry in the unitary operator as given in (2) and (20).



{\it Conclusions.--}
In this paper, we have shown the power of weak value amplification and postselection (WVAP) assisted quantum computation. However large the database may be, with the assistance of
 ancillary quantum register upon which we can perform pre- and post-selection, it is possible to find the desired item in one query. 
 It may happen that probability of postselection becomes very small for large database, but if it is non-zero, then in 
 that case the marked item can be searched in one step. We stress that in our model we do not allow postselection on the state of the quantum computer where the 
 algorithm is implemented.
 The exponential search algorithm comes at a cost, but that is bestowed by the ancillary quantum register.
 If we believe in the many world interpretation, then there is one branch of the ancillary register which will 
 certainly find the searched item in one query. 
 Since exponentially fast search algorithm can be conjured to solve NP problems, our result suggests that 
 it may be possible for quantum computers to solve NP problems with non-zero probability of success with WVAP quantum computer. 
 It will be interesting to implement the super quantum search algorithm 
 with the existing quantum computing platforms.

 Quantum computer assisted with a pre- and postselected quantum register can provide a new paradigm to discover quantum algorithms. At 
 the interpretation level, one can imagine that this algorithm is supplemented by ancillary quantum register  that is 
described by two state vectors and that act as a resource. The pre-selected state  propagates forward in
time and the postselected state propagates backward in time. Since two-time states are the basic objects in this
formalism they provide a boost to quantum search algorithm. 
 Instead of a sequence of unitary which involves $O(\sqrt{ N})$ steps, a 
single unitary interactions between the quantum computer and the ancillary register 
transforms the state of the quantum computer to the target state because of the existing correlations between forward and backward
states. The potent operator transforms the state of the quantum computer to the desired state in one query. 
The exponential speed-up that is possible here possibly arises from the fact that the ancillary quantum register conveys the answer about the target
state from the future.


\vskip 1cm
{\it Acknowledgement:} The possibility that weak value amplification with pre- and post-selections could speed-up quantum search algorithm 
was conjectured many years ago and the proof was completed in June 2018. I thank Samuel L. Braunstein, Aditi Sen De 
and Ujjwal Sen for several discussions. Also, I thank Sandu Popescu for useful remarks during 
a Workshop on Quantum Information and Quantum Foundations held in Crete, Greece during July 10-12, 2018.

\end{document}